\documentclass[%
 amsmath,amssymb,
 floatfix, 
 reprint,
 superscriptaddress,
 pra,
]{revtex4-2}

\usepackage{amsmath}
\usepackage{xcolor}
\usepackage{graphicx}
\graphicspath{ {./figs/} }
\usepackage{dcolumn}
\usepackage{bm}

\newcommand{\exb}[0]{\mathbf{E}\times\mathbf{B}}

\begin{document}

\title{Rapid cooling of the in-plane motion of two-dimensional ion crystals in a Penning trap to millikelvin temperatures}

\author{Wes Johnson}
\email{Wes.Johnson@colorado.edu}
\affiliation{Department of Physics, University of Colorado, Boulder, Colorado 80309, USA}%
\author{Athreya Shankar}
\affiliation{Department of Instrumentation and Applied Physics, Indian Institute of Science, Bangalore, India, 560012.}
\author{John Zaris} 
\affiliation{Department of Physics, University of Colorado, Boulder, Colorado 80309, USA}%
\author{John J. Bollinger}
\affiliation{National Institute of Standards and Technology Boulder}
\author{Scott E. Parker}
\altaffiliation[Also ]{Renewable and Sustainable Energy Institute, University of Colorado, Boulder}
\affiliation{Department of Physics, University of Colorado, Boulder, Colorado 80309, USA}

\date{\today}

\begin{abstract}

We propose a highly feasible technique with no experimental overhead to rapidly cool the in-plane degrees of freedom of large two-dimensional ion crystals in Penning traps. Through simulations, we demonstrate that our approach enables the in-plane modes to cool down to a temperature of around $1$ mK in less than $10$ ms.  Our technique relies on near-resonant coupling of the poorly cooled in-plane motions and the efficiently cooled out-of-plane motions, and is achieved without introducing additional potentials. The rapid cooling enabled by our approach is in contrast to typical operating conditions, where our simulations of the laser cooling dynamics suggest that the ion crystal's in-plane motion cools very slowly on a timescale of several hundreds of milliseconds, a rate likely slower than experimental heating rates. Our work sets the stage for sub-Doppler laser cooling of the planar motion, and more robust and versatile quantum simulation and quantum sensing experiments with two-dimensional crystals in Penning traps.

\end{abstract}

\maketitle

\begin{figure}
\includegraphics[width=\linewidth]{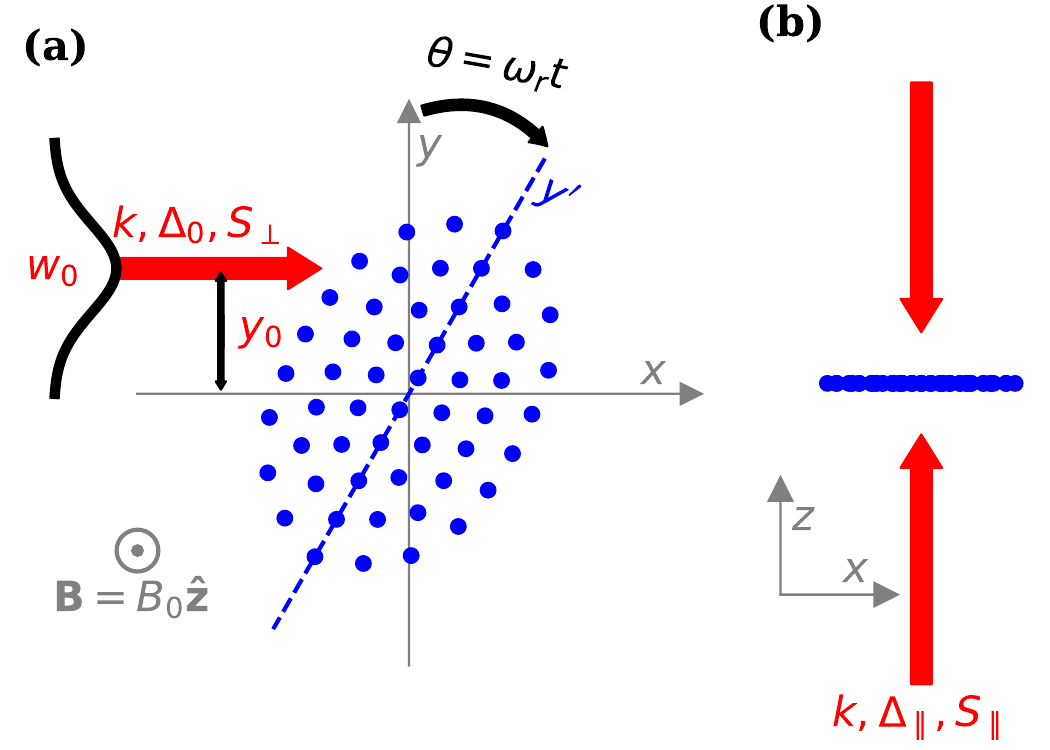} 
\caption{\label{fig:laserSetup} 
\textbf{Laser Cooling Setup.}  
\textbf{(a)} The planar laser beam. 
The lab frame coordinates are shown in grey, with the z axis chosen along the magnetic field, $B_0 = 4.458 \text{ T}$, and the x axis chosen parallel to the planar laser beam, depicted in red.  
The ion crystal is shown in blue.
A `rotating wall' potential precisely controls the rotation frequency, $\omega_r$, of the ion crystal. 
The simulation assumes $^9\text{Be}^+$ with a cooling transition wavelength $\lambda = 313 \text{ nm}$.
The planar beam is offset by $y_0 = 20 \text{ }\mu\text{m}$ from the center of the trap so that the laser's peak intensity occurs where the ions are receding from the laser source.   
The beam is assumed to have a Gaussian profile with a waist of $W_0 = 30 \text{ }\mu\text{m}$, and a saturation parameter $S_\perp = 1$. 
The saturation is given as a function of position, $S(y) = S_\perp \exp ( -2(y - y_0)^2/W_0^2)$.
The laser is red-detuned from resonance with the $^9\text{Be}^+$ laser cooling transition by $\Delta_0 = - 40 \text{ MHz}$.    
\textbf{(b)} The axial laser beam.
Two counter propagating beams are applied along the z axis to cool the drumhead modes.
Unlike in experiment, where only one beam is applied, two beams are used in simulation to cancel each other's displacement of the ion crystal's equilibrium, simplifying analysis. See SM for details. 
For these beams $S_\parallel = 5\times10^{-3}$ to achieve axial cooling with minimal recoil heating in the planar direction. 
The detuning is $\Delta_\parallel = - \gamma_0 /2$, where $\gamma_0 = 2\pi \times 18 \text{ MHz}$ is the natural linewidth of the $^9\text{Be}^+$ laser cooling transition -- this yields optimal cooling.   
}
\end{figure}

\begin{figure}
\includegraphics[width=\linewidth]{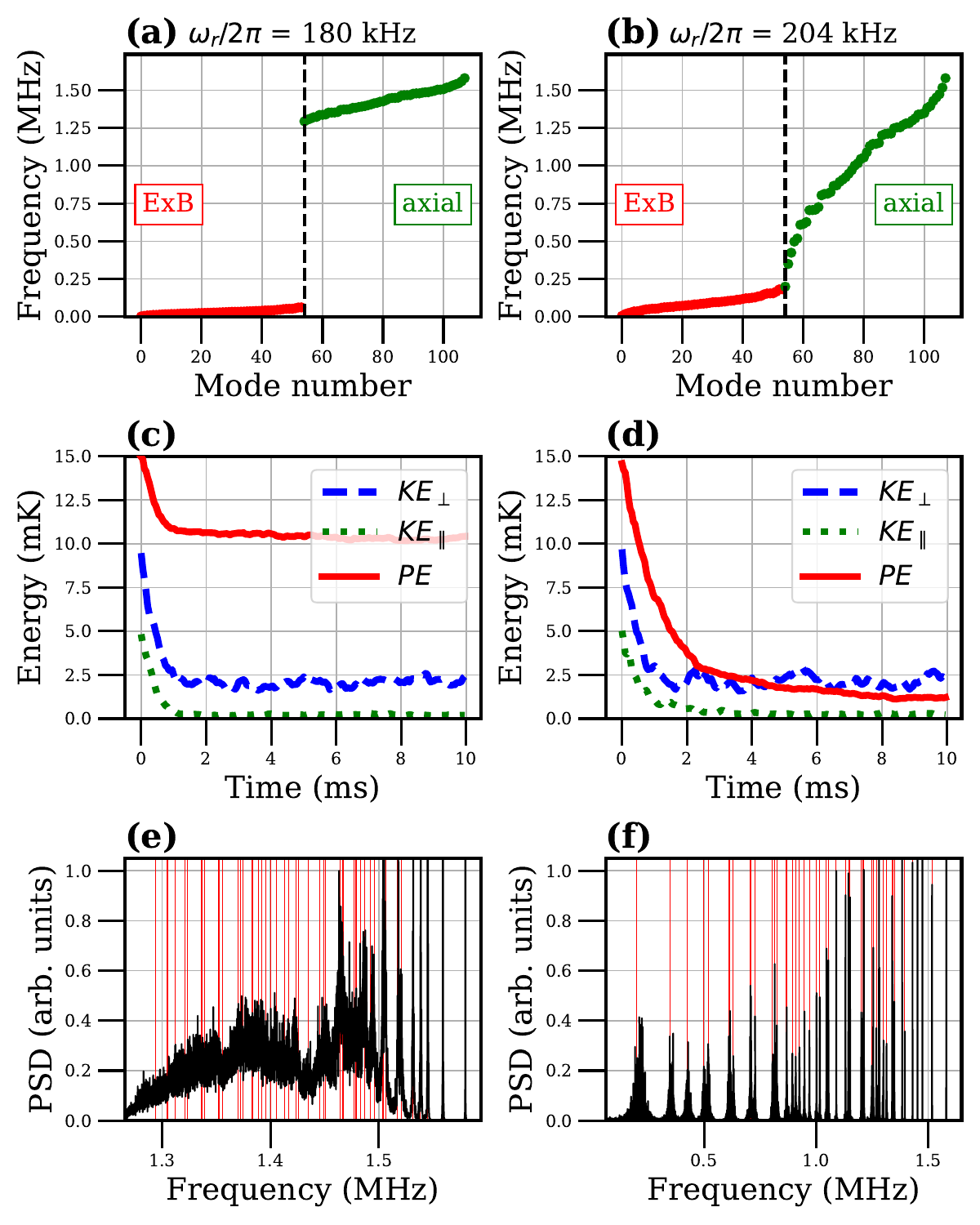}
\caption{\label{fig:coupling}
\textbf{Coupling of planar and drumhead modes via full simulation.}
A N = 54 ion crystal is initialized with all modes at an amplitude corresponding to a temperature of 10 mK.
These simulations are evolved with the full nonlinear Coulomb interaction and laser cooling for different values of the rotating wall frequency, $\omega_r$.    
\textbf{(a)} At $\omega_r / 2 \pi  = 180 \text{ kHz}$, the $\exb$ modes (left) are not resonant with the drumhead modes (right).   
\textbf{(b)} Increasing $\omega_r / 2 \pi $ to $204 \text{ kHz}$ brings the $\exb$ modes into resonance with the drumhead modes.    
\textbf{(c)} At $\omega_r / 2 \pi  = 180 \text{ kHz}$, the $\exb$ modes are not cooled substantially in 10 ms.   
\textbf{(d)} At $\omega_r / 2 \pi  = 204 \text{ kHz}$, the $\exb$ modes couple with the drumhead modes and are cooled to roughly 1 mK in 10 ms.     
\textbf{(e)} Drumhead mode spectra shows broadening due to uncooled $\exb$ modes.
\textbf{(f)} Drumhead mode spectra shows reduced broadening due to improved cooling of $\exb$ modes. 
}
\end{figure}

\begin{figure}
\includegraphics[width=\linewidth]{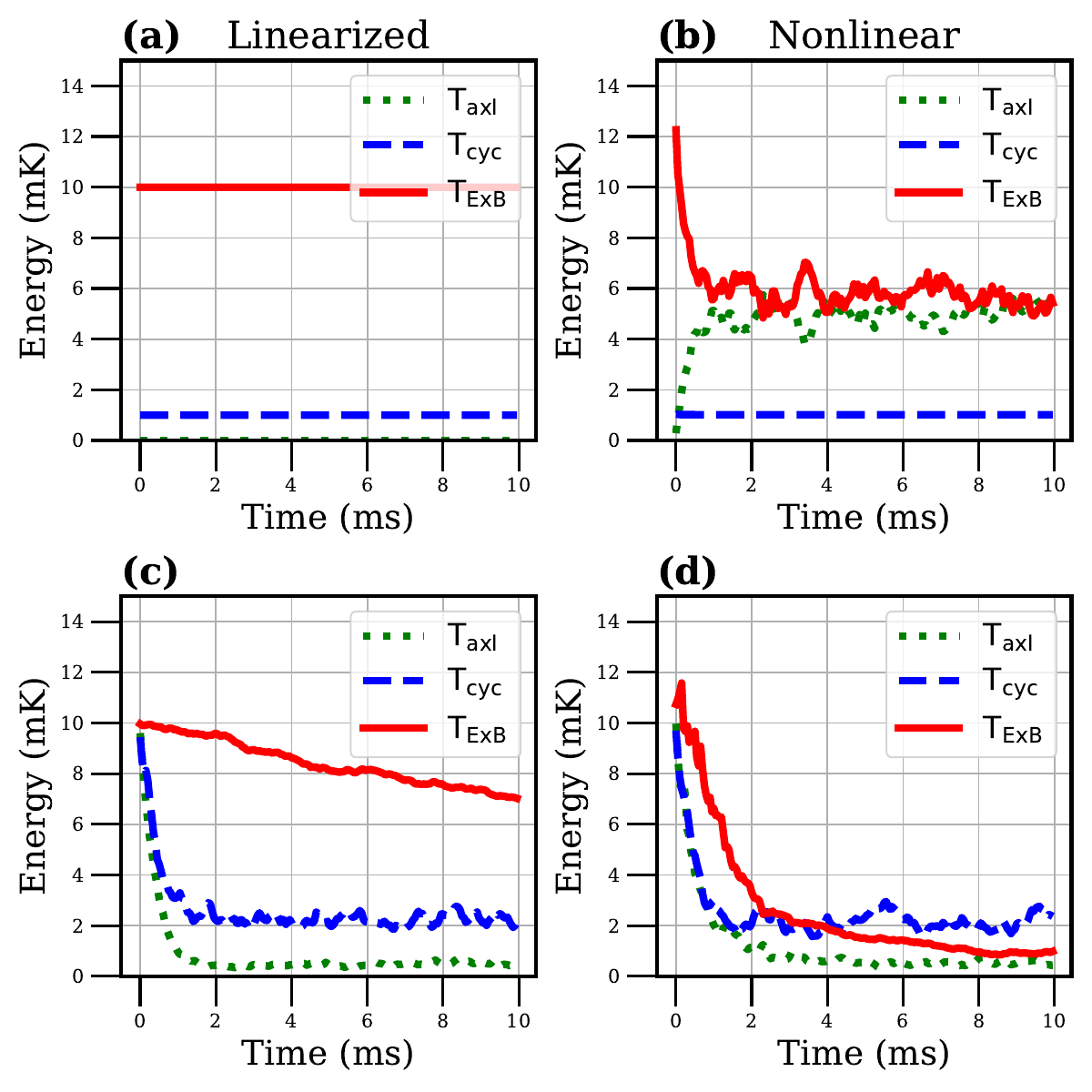}
\caption{\label{fig:Linearization}
\textbf{Comparison of linearized and full simulations.}
Linearized and full nonlinear simulations for a N = 54 ion crystal with $\frac{\delta}{\beta} = 0.25$ and $\omega_r / 2 \pi = 204 \text{ kHz}$ are evolved with and without laser cooling. 
\textbf{(a)} Linearized, without laser cooling, no change in mode branch temperatures observed.  
\textbf{(b)} Nonlinear, without laser cooling, rapid equilibration of $\exb$ and drumhead modes observed.  
\textbf{(c)} Linearized, with laser cooling, very modest reduction in $\exb$ mode temperature occurs.
\textbf{(d)} Nonlinear, with laser cooling, $\exb$ mode temperature reduced to roughly 1 mK.   
}
\end{figure}

\begin{figure*}
\includegraphics[width=\linewidth]{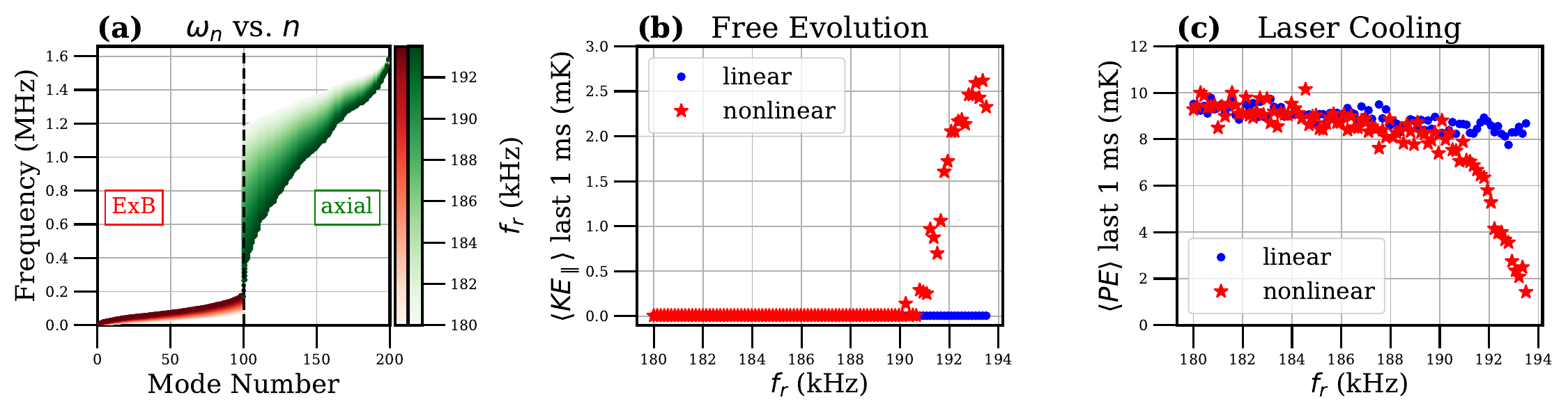}
\caption{\label{fig:fwallScan} 
\textbf{Rotating wall frequency scan.}
\textbf{(a)} The $\exb$ and drumhead mode frequencies vs. mode number colored by the rotating wall frequency for a N = 100 ion crystal.   
\textbf{(b)} The axial kinetic energy averaged over the last millisecond of evolution vs. rotating wall frequency for a N = 100 ion crystal evolved for 10 ms without laser cooling (free evolution). 
The simulations are initialized with $\exb$ mode branch temperatures of 10 mK, and zero drumhead and cyclotron mode branch temperatures.   
\textbf{(c)} The potential energy averaged over the last millisecond of evolution vs. rotating wall frequency for a N = 100 ion crystal evolved for 10 ms with laser cooling. 
The simulations are initialized with all mode amplitudes corresponding to a temperature of 10 mK.    
}
\end{figure*}

\emph{Introduction.}---Identifying routes to control systems with a large number of degrees of freedom is a crucial step in scaling up quantum technologies. Penning traps offer the ability to store and manipulate the electronic and motional states of a large number of ions at the quantum level~\cite{Monroe2021}. Several efforts are underway to utilize two-dimensional crystals of tens to several hundreds of ions stored in Penning traps for quantum sensing and quantum information processing~\cite{McMahon2020,McMahon2022,Ball2019,Wolf2023,Bohnet2016,Hrmo2019,Jain2020,Goodwin2015}. 
Researchers have designed protocols to simulate many-body quantum systems \cite{Britton2012,SafaviNaini2018,Cohn2018,Shankar2022,Qiao2022}, studied the spread of entanglement in interacting systems \cite{Grttner2017,Swingle2016}, and demonstrated spin squeezing and quantum-enhanced motion sensing protocols \cite{Bohnet2016,Gilmore2017,Affolter2020,Pezz2018,Gilmore2021,Toscano2006}. 
These protocols are enabled by coupling the electronic states of the ions to their out-of-plane normal modes of vibration--- called the drumhead modes--- using lasers. 
Although the in-plane normal modes of the ions are typically not utilized in these protocols, it was shown recently that low-frequency planar modes can significantly broaden the drumhead mode spectrum if they are not cooled well~\cite{Shankar2020}, limiting the utility of the drumhead modes as a quantum channel for mediating ion-ion interactions.

In this Letter, we numerically demonstrate that the Doppler laser cooling of the low-frequency planar modes can be greatly improved by resonantly enhancing their coupling to the drumhead mode branch, which itself is already efficiently laser cooled~\cite{Tang2019}. 
Through simulations, we show that this technique can cool the low-frequency planar modes to around $1$ mK in less than $10$ ms. 
In contrast, under typical experimental conditions \cite{Bohnet2016,Gilmore2021}, our simulations reveal that these modes are cooled very slowly on a timescale of hundreds of milliseconds, which is impractical in an actual experiment where extraneous heating effects will dominate. 
Furthermore, we demonstrate how the efficient cooling leads to significant improvement in the resolution of the drumhead mode spectrum. 
By comparing our results with a simulation that removes the coupling between modes, we elucidate the role played by the resonantly enhanced coupling between the planar and drumhead modes in improving the cooling.

\emph{Setup and Background.}---The setup we consider is shown in Fig.~\ref{fig:laserSetup}. Ions are confined in a Penning trap using a strong magnetic field $\mathbf{B}=B_0\mathbf{\hat{z}}$, and an electric quadrupole field generated by trap electrodes (not shown).  
Cooling lasers are applied along the $x$ direction and the $z$ direction to cool the planar ($\perp \mathbf{B}$) and axial ($\parallel \mathbf{B}$) motions respectively. 
The planar cooling laser is offset from the trap center. Furthermore, the crystal is rotating when viewed in the lab frame; the role of this rotation in the laser cooling will be discussed in more detail below. To make our work concrete, we choose trap and laser cooling parameters relevant to the NIST Penning trap~\cite{Bohnet2016}.  
Details of the simulation procedure are reported in the Supplemental Material (SM).

The nontrivial nature of the Doppler laser cooling in a Penning trap can be illustrated with a single trapped ion.  
The planar motion of a single ion is a superposition of cyclotron and magnetron modes~\cite{Brown1986,Kretzschmar1991}. 
Since the total energy of the magnetron mode is negative in the lab frame, reducing the amplitude of these motions requires that the laser must simultaneously add energy to the magnetron mode while removing energy from the cyclotron mode. This results in a fundamental cooling trade-off.  The amplitude of both motions can be reduced simultaneously with a laser beam tuned to the red of the atomic ion cooling transition and an intensity gradient applied across the center of the trap \cite{Itano1982,Thompson2000,Asprusten2013}.  The typical experimental method is to apply a focused planar ($\perp \mathbf{B}$) laser beam that is offset from the trap center, as is shown in Fig.~\ref{fig:laserSetup}, such that its peak intensity occurs where the magnetron motion is receding from the laser source.  Measured cooling times depend on the ion species and details of the cooling laser configuration, but in Refs. \cite{vanEijkelenborg1999,Phillips2008}, ranged from 10 to 200 ms for the magnetron motion. The measured cyclotron cooling was two to three orders of magnitude faster.

In the case of large multi-ion crystals with tens to hundreds of ions, a complete theoretical treatment of the planar laser cooling of the ion crystal is challenging. To investigate this system, we therefore perform numerical simulations using a full-dynamics integrator that includes a realistic laser cooling model~\cite{Tang2019}.

In contrast to a single ion, Doppler laser cooling of multi-ion crystals in a Penning trap is complicated by the collective rotation of the crystal as viewed in the lab frame (see Fig.~\ref{fig:laserSetup}). 
This leads to coherent Doppler shifts across the ion crystal and further limits the lowest attainable perpendicular kinetic energy ($\mathrm{KE}_\perp$)~\cite{Itano1988}. Additionally, the focused laser beam generates a torque on the ion crystal, and can change the collective rotation frequency, $\omega_r$, of the ion crystal~\cite{Bollinger1995,Torrisi2016,Asprusten2013}. 
A rotating quadrupolar potential, called a `rotating wall,' is applied to precisely control $\omega_r$ in experiments~\cite{Huang1998}. 
Energy exchange with the rotating wall potential allows for cooling of $\mathrm{KE}_\perp$ to millikelvin temperatures~\cite{Torrisi2016}. We chose the offset $y_0$ and the laser detuning $\Delta_0$, shown in Fig.~\ref{fig:laserSetup}, to roughly minimize the laser torque for the beam width of $W_0 = 30 \text{ }\mu\text{m}$, and to maximize the cooling of $\mathrm{KE}_\perp$ as calculated by a simple model presented in Ref.~\cite{Torrisi2016}.   

However, efficient cooling of $\mathrm{KE}_\perp$ does not necessarily imply effective cooling of the potential energy fluctuations ($\mathrm{PE}$) that are associated with planar motions.
The presence of a strong trapping magnetic field $\mathbf{B}$ leads to unconventional normal modes in the planar direction, which can be classified into a low-frequency $\exb$ branch dominated by $\mathrm{PE}$ and a high-frequency cyclotron branch dominated by $\mathrm{KE}_\perp$~\cite{Shankar2020,Dubin2020}. 
Simulations suggest that the $\exb$ and cyclotron modes do not equilibrate~\cite{Tang2021}, allowing the possibility for large $\mathrm{PE}$ to persist despite efficient cooling of $\mathrm{KE}_\perp$. 
This is in contrast to the drumhead modes, whose total energy is, on average, equally shared between axial kinetic energy ($\mathrm{KE}_\parallel$) and $\mathrm{PE}$, and hence cooling $\mathrm{KE}_\parallel$ leads to an equal reduction of the $\mathrm{PE}$. From our simulations, we find that the $\mathrm{PE}$ associated with planar motion is not efficiently cooled under current experimental conditions. Interestingly, as shown in Fig.~\ref{fig:coupling}(c), our simulations suggest the potential and kinetic energies have vastly different cooling rates.
$\mathrm{KE}_\perp$ and $\mathrm{KE}_\parallel$ are cooled to millikelvin temperatures in roughly $1$ ms, whereas after a brief cooling related to the reduction in $\mathrm{KE}_\parallel$, $\mathrm{PE}$ is not significantly cooled.

\emph{Resonant Mode Coupling.}---In order to efficiently cool the $\exb$ modes, we investigate a method for sympathetically cooling these modes by resonantly coupling them to the drumhead mode branch, which itself is well-cooled. 
For experiments utilizing a rotating wall potential, our technique is significant in that it requires no additional time-dependent potentials to engineer couplings \cite{Brown1986,Hou2023,Hendricks2008}, nor a change to current laser cooling setups.
Instead, it relies on the observation that the frequency gap between the drumhead and $\exb$ mode branches can be tuned by changing the ion crystal rotation frequency $\omega_r$, which is a precisely controlled experimental parameter. 
In particular, this gap can be closed through an appropriate choice of $\omega_r$, which leads to a resonant enhancement of the inter-branch coupling. 

To understand this, we consider the potential energy of the ions in a non-inertial reference frame rotating with the crystal, where the total Lagrangian is time independent. This potential energy can be written as  
\begin{equation}
\label{eqn:potential}
\begin{aligned}
U_r= & \sum_{i=1}^N \frac{1}{2} m \omega_z^2 \left( z_i^2
+ \left( \beta + \delta \right) x_i^2 + \left( \beta - \delta \right) y_i^2 \right) \\
& +\sum_{i=1}^N \sum_{j \neq i} \frac{q^2}{8 \pi \varepsilon_0} \frac{1}{\left|\mathbf{x}_i-\mathbf{x}_j\right|}.
\end{aligned}
\end{equation}
Here, $\beta$ characterizes the relative strength of planar and axial confinement and is given by
\begin{equation}
\beta = \frac{\omega_r\left(\omega_c-\omega_r\right)}{\omega_z^2} - \frac{1}{2},    
\end{equation}
where $\omega_c$ is the cyclotron frequency and $\omega_z$ is the axial trap frequency. 
The azimuthal asymmetry due to the rotating wall potential, which is static in this co-rotating frame, is parameterized by $\delta$.
In the rotating frame, the potential energy of the ions is always positive. 
Potential energy fluctuations are calculated as $\text{PE} = \Delta U_r$, the increase in potential energy from the the crystal's equilibrium configuration. 
The value of $\beta$ can be increased by increasing the rotating wall frequency. At a critical value $\omega_{r,\text{crit.}}$, where $\beta=\beta_\text{crit}$, the planar configuration becomes unstable, and the ion crystal transitions to a 3D configuration~\cite{Dubin1993}.    
Near $\beta_\text{crit.}$ the lowest frequency drumhead mode approaches zero, which represents an instability of the planar ion crystal to axial displacements.  
In our simulations, we increase $\omega_r$ to near, but less than, $\omega_{\text{crit.}}$, such that the lowest frequency drumhead modes are nearly resonant with the highest frequency $\exb$ modes.

\emph{Results.}---In Fig.~\ref{fig:coupling}, we show the results of simulations for an $N=54$ ion crystal at a rotating wall frequency of $\omega_r/(2\pi) = 180 \text{ kHz}$, typical of NIST work~\cite{Bohnet2016, Gilmore2021}, and an increased rotating wall frequency of $\omega_r/(2\pi) = 204 \text{ kHz}$. 
In Fig.~\ref{fig:coupling}(a) and \ref{fig:coupling}(b), the $\exb$ and drumhead mode frequencies are plotted as a function of mode number for the ion crystal simulated at these two values of $\omega_r$. 
By increasing $\omega_r$, the lowest frequency drumhead  modes are brought into resonance with the $\exb$ modes.  
Previous numerical studies have shown that mode coupling between $\exb$ modes leads to rapid equilibration of these modes \cite{Tang2021}, thus suggesting that coupling only a few $\exb$ modes to drumhead modes may be sufficient to sympathetically cool the entire $\exb$ mode branch. 
In Fig.~\ref{fig:coupling}(c) and \ref{fig:coupling}(d), $\mathrm{KE}_\parallel$, $\mathrm{KE}_\perp$,and $\mathrm{PE}$ are plotted as a function of time for the two cases.
These energies are calculated directly from the ion positions and velocities during the evolution of the simulations.  
The energies are normalized to the ion number and then converted to temperature units via $T = E/(Nk_B)$, where $E\in\{\mathrm{KE}_\parallel,\mathrm{KE}_\perp,\mathrm{PE}\}$, and $k_B$ is the Boltzmann constant. 
In these simulations, we initialize the ion crystal with all mode amplitudes corresponding to 10 mK, then a random phase is chosen for each mode.
Details are given in SM.
Cooling of the kinetic energies is similar in both simulations; however, the cooling of the potential energy is much faster when $\omega_r/(2\pi)=204 \;\text{kHz}$.  
From longer simulations at $\omega_r/(2\pi)=180\;\text{kHz}$, we found that $\mathrm{PE}$ does cool, however, this cooling process takes hundreds of milliseconds (see SM).    
In contrast, Fig.~\ref{fig:coupling}(d) shows that $\mathrm{PE}$ is cooled to roughly 1 mK after 10 ms of laser cooling when $\omega_r/(2\pi)=204 \;\text{kHz}$. 

The improved cooling of the $\exb$ modes, i.e., the $\mathrm{PE}$ associated with planar motion, leads to reduced fluctuations in ion positions. 
Consequently, the adverse impact of planar position fluctuations on the drumhead mode spectrum is greatly reduced. 
In Fig.~\ref{fig:coupling}(e) and \ref{fig:coupling}(f), we plot the power spectrum of the drumhead motion for the two cases considered here, after the crystals have been laser cooled for $10$ ms. For $\omega_r/(2\pi)=180 \;\text{kHz}$ [Fig.~\ref{fig:coupling}(e)], the drumhead modes are so strongly broadened that the spectrum appears as a smooth continuum beyond the first few well-resolved highest frequency modes. 
In general, the sensitivity of the drumhead modes to planar position fluctuations is stronger for the lower frequency modes. 
The spectrum shown in Fig.~\ref{fig:coupling}(e)  is consistent with Fig.~(7c) of Ref.~\cite{Shankar2020}, which shows the broadening of the drumhead spectrum due to an in-plane temperature of 10 mK. 
In contrast, when $\omega_r/(2\pi)=204\;\text{kHz}$ [Fig.~\ref{fig:coupling}(f)], the drumhead spectrum shows resolved peaks near the predicted drumhead mode frequencies over a much larger range of mode frequencies. 
Improving the drumhead spectral resolution could enable the use of more of these modes for high fidelity quantum information processing protocols. 
So far, protocols have primarily utilized only the center-of-mass mode, which is the highest frequency drumhead mode and is insensitive to planar position fluctuations. 

To understand the role played by the mode coupling in the improved cooling, we compared our results against a simulation where the Coulomb interaction, given by the last term in Eq.~(\ref{eqn:potential}), was expanded to second order in the displacements of the ions from their equilibrium positions. 
The expansion results in a Coulomb force that is linear in the displacements of the ions. 
This ``linearization" removes the mode coupling due to higher order terms, and allows for the direct laser cooling to be isolated in simulations. 
The details of the linearization procedure and the associated simulations are provided in the SM.

Fig.~\ref{fig:Linearization} compares the results of the linearized and the full-dynamics (nonlinear) simulations for an $N = 54$ ion crystal at a rotating wall frequency of $\omega_r/(2\pi) = 204 \text{ kHz}$ in the presence and absence of laser cooling.
Furthermore, the mode branch temperatures in Fig.~\ref{fig:Linearization} are calculated directly from the amplitudes of the modes during evolution. 
These mode energy averages, neglecting nonlinear corrections from the Coulomb potential, provide an approximation of the system's energy at low temperatures. 
The mode initialization procedure and calculation of the mode branch temperatures are discussed in Ref.~\cite{Tang2021} and summarized in the SM.

In Figs.~\ref{fig:Linearization}(a) and \ref{fig:Linearization}(b), we investigate the free evolution of the crystal in the absence of the cooling lasers.  
The ion crystal was initialized with $\exb$ modes at 10 mK, cyclotron modes at 1 mK, and drumhead modes at 0 mK.
In Fig.~\ref{fig:Linearization}(a), mode branch temperatures are unchanged over $10$ ms of free evolution, since the linearization of the dynamics removes the mode coupling and energy cannot be exchanged between the different branches. 
In Fig.~\ref{fig:Linearization}(b), the equilibration of the $\exb$ and drumhead modes occurs rapidly, in roughly 1 ms, demonstrating the strong mode coupling arising from the full nonlinear evolution. 
We note that the ExB temperatures plotted in \ref{fig:Linearization}(b) and \ref{fig:Linearization}(d) exceed 10 mK due to nonlinear corrections neglected in the mode energy calculation, particularly significant for low-frequency ExB modes at energies around 10 mK.

In Figs.~\ref{fig:Linearization}(c) and \ref{fig:Linearization}(d), the linearized and nonlinear simulations were integrated with laser cooling. 
The nonlinear simulation shown in Fig.~\ref{fig:Linearization}(d) is the same as the one shown in Fig.~\ref{fig:coupling}(d), although in Fig.~\ref{fig:Linearization}(d) mode branch temperatures are plotted instead of $\mathrm{KE}_\parallel$, $\mathrm{KE}_\perp$, and $\mathrm{PE}$. 
In the linearized case, although there is a reduction in the $\exb$ mode temperature during the $10$ ms of laser cooling, this temperature is still many times larger than the cyclotron and drumhead mode temperatures. 
In contrast, in the nonlinear simulation, the $\exb$ mode temperature is rapidly reduced to roughly 1 mK after $10$ ms of laser cooling. 
The difference between the linear and nonlinear simulations further illustrates that coupling between the $\exb$ and drumhead modes due to the nonlinear Coulomb interaction is responsible for the accelerated cooling of the $\exb$ modes.

So far, we compared the cooling dynamics at two representative values of $\omega_r$ and demonstrated the role of the mode coupling when $\omega_r$ is close to $\omega_\text{crit.}$. To further investigate the effect of mode coupling on the cooling of the $\exb$ modes, we studied the cooling dynamics and mode coupling as  $\omega_r$ was scanned. In Fig.~\ref{fig:fwallScan}, an $N = 100$ ion crystal was initialized and evolved for a range of rotating wall frequencies between $\omega_r/(2\pi) = 180 \text{ kHz}$ and $\omega_r/(2\pi) = 194 \text{ kHz}$. The larger number of ions lowers the critical rotating wall frequency to roughly $\omega_{\text{crit.}}/(2\pi) = 194.75 \text{ kHz}$.

In Fig.~\ref{fig:fwallScan}(a), the drumhead and $\exb$ mode frequencies are plotted as $\omega_r$ is scanned, with the color gradient specifying the $\omega_r$ value. Notably, the lower frequency drumhead modes are rapidly brought into resonance with the higher end of the $\exb$ branch as $\omega_r$ approaches within a few kHz of $\omega_{\text{crit.}}$. 
The $\exb$ mode frequencies, however, only change slightly as $\omega_r$ is increased.

In Fig.~\ref{fig:fwallScan}(b), we study the energy exchange between the $\exb$ and drumhead mode branches when the crystal is freely evolved in the absence of cooling lasers for $10$ ms. Here, the $\exb$ mode branch is initialized with a temperature of 10 mK while the other modes are initialized to 0 mK. We plot the average $\mathrm{KE}_\parallel$ during the last $1$ ms of the simulation as a function of $\omega_r$. The linearized simulation shows no change in $\mathrm{KE}_\parallel$, whereas the nonlinear simulation shows an increase in the average $\mathrm{KE}_\parallel$ as $\omega_r$ is increased. Near  $\omega_{\text{crit.}}$, $\mathrm{KE}_\parallel$ increases to roughly $2.5$ mK, which is consistent with the drumhead branch temperature being roughly $5$ mK. 
This suggests that the $\exb$ and drumhead branches equilibrate within $10$ ms for $\omega_r$ close to $\omega_{\text{crit.}}$. However, for lower $\omega_r$, $\mathrm{KE}_\parallel$ in both the linear and nonlinear simulations is unchanged, indicating that the coupling between branches only becomes significant near $\omega_{\text{crit.}}$.

In Fig.~\ref{fig:fwallScan}(c), the simulation was integrated with laser cooling for 10 ms.
The average total potential energy during the last 1 ms of the simulation is plotted as a function of $\omega_r$. The linearized simulation shows a modest reduction in $\mathrm{PE}$ as the $\omega_r$ is increased, which may be due to the increased fraction of $\mathrm{KE}_\perp$ contributing to the $\exb$ mode energies.
However, in the nonlinear simulation, $\mathrm{PE}$ is rapidly reduced as $\omega_r$ approaches $\omega_{\text{crit.}}$. Near $\omega_{\text{crit.}}$, the average $\mathrm{PE}$ over the last $1$ ms is roughly 1 mK.

\emph{Conclusion and Outlook.}---We have demonstrated a technique to efficiently cool the low-frequency planar motion of large 2D ion crystals in Penning traps, which has hitherto been challenging. 
Our technique has no experimental overhead and can be used to rapidly initialize ($< 10$ ms) crystals with all motional degrees of freedom cooled down to millikelvin temperatures.
This achievement sets the stage for for sub-Doppler limit laser cooling studies encompassing all $3N$ motional modes. 
Furthermore, we showed how the improved planar cooling greatly reduced the spectral broadening of the drumhead modes. 
The improved resolution of the drumhead modes expands the scope of quantum information protocols that can be performed with large 2D crystals stored in Penning traps. As a result, it is immediately relevant for several experiments aiming to use Penning traps for quantum information processing~\cite{McMahon2020,Ball2019,Bohnet2016}. 

We note that overlapping the bandwidth of the drumhead modes with the $\exb$ modes introduces low-frequency drumhead modes. 
For a given temperature of the drumhead motion this results in correspondingly larger Lamb-Dicke confinement parameters, which can impact the fidelity of quantum operations \cite{Wineland1998,Srensen2000}.
Ground-state cooling of the drumhead modes can help \cite{Jordan2019,Shankar2019,Kiesenhofer2023}, but experimental constraints may dictate a lower bound on the drumhead mode frequencies that is significantly higher than any of the $\exb$ modes. 
In this case it may be possible to use appropriate time-dependent electric field configurations to couple the $\exb$ modes to either the drumhead modes or the cyclotron modes, which are efficiently laser cooled. 
Techniques like this have been investigated with single and small numbers of trapped ions \cite{Brown1986,Hou2023}. 
A notable example is the so-called axialization technique, which reduces the size of the magnetron motion of a small number of ions in a Penning trap by coupling the magnetron motion with the cyclotron motion \cite{Powell2002,Hendricks2008,Hrmo2019}. 
Alternately, there is the possibility of adiabatically decreasing $\omega_r$ or similarly increasing $\omega_z$ after cooling the $\exb$ modes, which not only eliminates the nonlinear coupling introduced, but would also increase the frequency of the lowest frequency drumhead modes. 
These possibilities motivate interesting directions for future numerical simulations.

\begin{acknowledgments}
We thank Bryce Bullock for helpful discussions, Chen Tang for guidance and support in the initial stages of this project, and Jennifer Lilieholm and Mason Marshal for comments on the manuscript. Work supported by U.S. Department of Energy under grant number S0154230-7. A. S. acknowledges the support of a C. V. Raman Post-Doctoral Fellowship, IISc. JJB acknowledges support from DOE, Office of Science, Quantum Systems Accelerator, from AFOSR, and from the DARPA ONISQ program.
\end{acknowledgments}

\bibliography{bib}{}
\bibliographystyle{apsrev4-2}

\end{document}


\title{Supplemental Material}

\author{Wes Johnson}
\email{Wes.Johnson@colorado.edu}
\affiliation{Department of Physics, University of Colorado, Boulder, Colorado 80309, USA}%
\author{Athreya Shankar}
\affiliation{Department of Instrumentation and Applied Physics, Indian Institute of Science, Bangalore, India, 560012.}
\author{John Zaris} 
\affiliation{Department of Physics, University of Colorado, Boulder, Colorado 80309, USA}%
\author{John Bollinger}
\affiliation{National Institute of Standards and Technology Boulder}
\author{Scott E. Parker}
\altaffiliation[Also ]{Renewable and Sustainable Energy Institute, University of Colorado, Boulder}
\affiliation{Department of Physics, University of Colorado, Boulder, Colorado 80309, USA}

\date{\today}

\maketitle

\section{NIST Penning Trap Parameters and Simulation}

\begin{figure}[h!]
    \includegraphics[width=\linewidth]{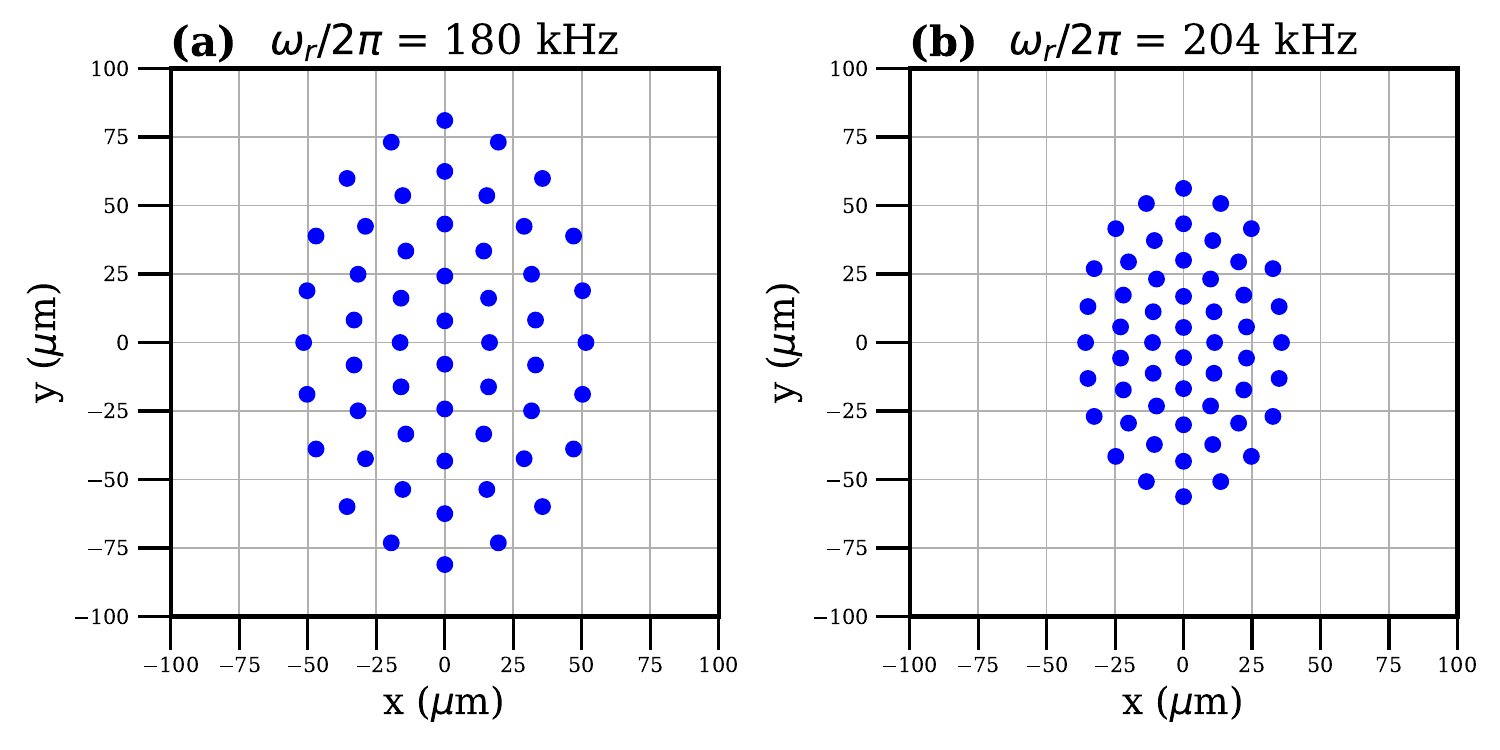} 
    \caption{\label{fig:crystal} 
    \textbf{N = 54 Ion Crystal Equilibria.}
    Two planar ion crystal equilibrium configurations corresponding to different rotating wall frequencies are shown in their respective rotating frames. 
    The ratio of $\delta/\beta = 0.25$ is fixed for both crystals, such that the aspect ratios of the ion crystals being studied are the same.
    \textbf{(a)} With $\omega_r = 2\pi \times 180 \text{ kHz}$ the ion crystal is less dense. 
    \textbf{(b)} With $\omega_r = 2\pi \times 204 \text{ kHz}$ the ion crystal is noticeably smaller. 
    }
\end{figure}   

Our study uses a molecular dynamics-like simulation to study laser cooling of large ion crystals in the Penning trap. 
The ions are treated as point particles with positions $\mathbf{r}_i$ and velocities $\mathbf{v}_i$ for $i = 1,2,\dots N$, where $N$ is the number of ions. 
The equations of motion are evolved numerically in the lab frame using a time splitting algorithm equivalent to the Buneman and Boris algorithms. 
The laser interaction is modeled stochastically by considering resonance fluorescence of a simplified two-level atom. 
For each timestep and for each ion, the mean number of photons scattered is calculated, $\bar n = \dot n \Delta t$, where $\dot n$ is the scattering rate. 
For our choice of timestep, $\Delta t = 10^{-9}$, $\bar n \ll 1$, meaning scattering of multiple photons by a single ion in one timestep is rare. 
The number of photons scattered is generated from a Poisson distribution with mean $\bar n$. 
The recoil from re-emission is assumed to be isotropic. 
The velocity of the ion is updated based on the simulated net momentum kick from absorption and re-emission.
Although $\Delta t = 10^{-9}$ for all simulation results presented in this study, we only save positions and velocities once every 1000 time steps, or each microsecond. 
For the plotted energy values versus time, rather than being plotted every microsecond, each point represents the average of 100 consecutive energy calculations. 
Details of our simulation framework are described in Ref.~\cite{Tang2019}.

One example of planar ion crystals used in Penning trap experiments is from the Ion Storage Group at the National Institute of Standards and Technology (NIST), which routinely manipulates planar ion crystals consisting of hundreds of ions in quantum sensing and simulation experiments \cite{Affolter2020,Ge2019,Bohnet2016}. 
Our simulation parameters are chosen to reflect those of the NIST experiment. 
The NIST Penning trap magnet is a superconducting high-field magnet with a strength near the trap center of roughly $4.46\text{ T}$. 
Typically, $^9\text{Be}^+$ ions are trapped in this device.
The $^9\text{Be}^+$ ions are an attractive ion species for quantum protocols. 
Their hyperfine Zeeman levels can serve as the qubit in quantum information experiments, and their $^2S_{1/2} \rightarrow ^2P_{3/2}$ transition allows for an efficient Doppler laser cooling cycle. 
The bare cyclotron frequency of a single $^9\text{Be}^+$ ion in the NIST trap is $\omega_c \approx 2\pi\times 7.60\text{ MHz}$, reflected by the magnetic field strength of $B = 4.4588\text{ T}$ chosen for simulations. 
The axial trap frequency is $\omega_z = 2\pi\times 1.58 \text{ MHz}$, chosen from experiments.
To control the collective rotation frequency of the $^9\text{Be}^+$ ion crystal, a rotating wall potential is applied at a precisely controlled frequency. 
In recent NIST Penning trap experiments \cite{Gilmore2021,Bohnet2016} a rotating wall frequency of roughly $\omega_r = 2\pi \times 180\text{ kHz}$ has been applied. 
In simulations, we consider a range of $\omega_r$ that is accessible to the NIST Penning trap experiment.

The ions stored in the NIST Penning trap are laser cooled via a combination of axial and planar laser beams. 
The $^9\text{Be}^+$ laser cooling transition has a wavelength of $\lambda = 313\text{ nm}$ and natural linewidth $\gamma_0 = 2\pi\times 18 \text{ MHz}$, which correspond to the $^2S_{1/2} \rightarrow ^2P_{3/2}$ transition.        
The axial beam width is much larger than the planar extent of the ion crystal and its intensity can be considered uniform for the purposes of our simulations. 
Only one axial laser beam is used in the experiment, but in our simulations we use two counter-propagating axial laser beams. 
This ensures that there is no unbalanced force shifting the ion crystal's equilibrium position in simulation. 
In the experiment, the axial laser beam is adiabatically turned off to avoid exciting the axial center of mass motion as the radiation pressure from the axial laser beam is removed.
The laser cooling dynamics proceeds similarly in both cases, however, the two beams in the simulation simplify the analysis of the ion crystal's motion. 
These beams have a weak saturation, chosen in our simulations to be $S_z = 5 \times 10 ^{-3}$. 
The axial laser beams have a detuning from resonance with the laser cooling transition of $^9\text{Be}^+$ equal to $\Delta_z = - \gamma_0/2$. 
This red detuning is chosen to minimize the Doppler cooling limit of a single ion's axial motion \cite{Itano1982}.   
In the NIST experiment, Doppler laser cooling of axial modes can achieve temperatures near the Doppler cooling limit of $T_D = 0.43 \text{ mK}$ \cite{Sawyer2014}.

Laser cooling of ion clouds in a Penning trap can generate crystalline structures \cite{Tan1995}.  
Although the ions can never be motionless due to the velocity dependent nature of the confining magnetic force, the rotating frame, defined by the collective rotation of the ion crystal and controlled by the rotating wall potential, offers a unique reference frame for describing the ion crystal. 
After laser cooling the ions' motion, the ions self-organize into a two dimensional triangular crystalline lattice in the rotating frame \cite{Wang2013}. 
By minimizing their external trapping and Coulombic potential energies, given in Eq.~(1) of the main text,  we can calculate equilibrium positions of the ions. 
These equilibria are depicted in Fig.~\ref{fig:crystal} for the case of N = 54 ions at two different values of $\omega_r$. 
The two rotating wall frequencies chosen are $\omega_r/(2\pi) = 180 \text{ kHz}$ for Fig.~\ref{fig:crystal}(a) and $\omega_r/(2\pi) = 204 \text{ kHz}$ for Fig.~\ref{fig:crystal}(b). 
Note that the ion crystal in Fig.~\ref{fig:crystal}(b) is smaller than the one in Fig.~\ref{fig:crystal}(a) due to an increase in the velocity dependent planar confinement parameter $\beta$.
The rotation frequency of $\omega_r/(2\pi) = 204 \text{ kHz}$ is near $\omega_{\text{crit.}}$, the rotation frequency at which the planar ion crystal transitions to a 3D ion crystal. 
The value of $\omega_{\text{crit.}}$ for an N = 54 ion is roughly $\omega_r/(2\pi) \approx 204.57 \text{ kHz}$.
The transition arises from the increased planar confinement parameter $\beta$ as $\omega_r$ increases. 
An ion crystal with a three dimensional equilibrium configuration is generated beyond the critical rotating wall frequency $\omega_{\text{crit.}}$. 
The value of $\omega_{\text{crit.}}$ can be calculated numerically.
The calculation can be done by minimizing the potential energy in the rotating frame for different values of $\omega_r$ until a transition from a planar to 3D equilibrium is observed. 
The value of $\omega_{\text{crit.}}$ decreases as the number of ions in the crystal increases, and may also change with different rotating wall strengths. 

The planar ion crystals in Fig.~\ref{fig:crystal} have roughly the same aspect ratio. 
This is achieved by fixing the ratio of the anisotropy from the rotating wall potential (rotating wall strength), $\delta$, and the planar confinement parameter, $\beta$. 
In our simulations, $\delta/\beta = 0 .25$. 
For the N = 54 ion crystal, it has been empirically observed in simulations that this choice of parameters gives robustness against reorganization events in the plane. 
The robustness may be attributed to the apparent bilateral symmetry for the crystal shown in Fig.~\ref{fig:crystal}, since reorganization events have commonly been observed to be transitions between degenerate potential energy configurations that are reflections of each other about the $x\text{ or }y$ axis. 
In the absence of reorganization, the planar mode amplitudes during the evolution of the system can be calculated from displacements from equilibrium, see Fig.~3 of the main text. 
In general, however, planar reorganization events prevent the calculation of mode amplitudes from displacements, and we default to calculating $\text{PE }, \text{KE}_\perp, \text{ and } \text{KE}_\parallel$ during the evolution of the system. 
A notable exception is the linearized simulation, which removes the possibility of the reorganization events.

\section{Linearization}

To isolate the effect of mode coupling between the axial and $\exb$ mode branches from the laser cooling dynamics, we implemented a linearized simulation of the ion crystal's evolution. 
This consists of calculating the linear force response of the ions to displacements from equilibrium, and using this force to update the velocities and positions of the ions.
Since the external trapping potentials are quadratic in the displacements of ions from equilibrium, the Coulomb force is the only source of non-linearity in the system's evolution. 
Therefore, all other force calculations remain the same in the linearized simulation, save the linearized Coulomb force calculation discussed below. 
The Coulomb potential is given in Eq.~(\ref{eqn:coulomb}) below:

\begin{equation}
\label{eqn:coulomb}
U_\text{C} = \frac{e^2}{8\pi \epsilon_0}\sum_{i=1}^N \sum_{j \neq i} \frac{1}{\left|\mathbf{x}_i-\mathbf{x}_j\right|}. 
\end{equation}
The expansion up to second order in terms of displacements from the equilibrium positions $x_i^0, \quad i = 1,2,\ldots,3N$ is given by 
\begin{equation}
\label{eqn:expansion}
\begin{aligned}
U_\text{C} = U_\text{C,0} + \sum_{i=1}^{3N} J_i q_i + \frac{1}{2}\sum_{i,j}^{3N} H_{ij} q_i q_j, \quad \mathbf{q} = \mathbf{x} - \mathbf{x}^0, \\
J_i = \frac{\partial U_C}{\partial x_i} \bigg|_{\mathbf{x}^0}, \quad   
H_{ij} = \frac{\partial^2 U_C}{\partial x_i \partial x_j} \bigg|_{\mathbf{x}^0},
\end{aligned}
\end{equation}  
where $U_\text{C,0}$ is the Coulomb potential energy at equilibrium, and $J_i$ and $H_{ij}$ are respectively the Jacobian and Hessian of the Coulomb potential energy evaluated at the equilibrium configuration.
Note that at equilibrium, $J_i$ is exactly canceled by the the Jacobian arising from the external trapping potential. 
The dynamical matrix for the system can be constructed from the second derivatives of the total potential (external trapping and Coulomb repulsion), and describes the linear evolution of the system's eigenmodes \cite{Shankar2020,Dubin2020}. 
The Coulomb potential is the only component of the total potential with nonzero third order and higher derivatives.
Higher order derivatives encapsulate nonlinear processes that will change the mode amplitudes over time, such as mode coupling for example, which can allow for energy redistribution between modes. 
Mode coupling between the planar modes has been studied numerically previously \cite{Tang2021}.
Here however, we calculate the linear force response of the ion crystal due to small displacements from equilibrium, which will neglect nonlinear effects. 
To calculate the linearized force on the ions, we take the gradient of the potential energy in Eq.~(\ref{eqn:expansion}) with respect to displacements $q_i$. 
The components of the linearized Coulomb force on the ions are given by

\begin{equation}    
\label{eqn:linearizedforce}
\begin{aligned}
F_i = -J_i - \sum_{j=1}^{3N} H_{ij} q_j
\end{aligned}
\end{equation}

To calculate this force and update the linearized simulation, we first rotate the lab frame coordinates into the rotating frame, then calculate the displacements from equilibrium, $q_i$.  
Next, we calculate the components of the linearized Coulomb force acting on the ions using Eq.~(\ref{eqn:linearizedforce}), and rotate this force back into the lab frame.  
The coordinate rotations can be done using the rotation matrix in Eq.~(\ref{eqn:rotatingmatrix}), and changing the sign of $\omega_r$ correspondingly.
Finally, we update the velocities and positions of the ions using the linearized force in place of the full Coulomb force, whereas all other force calculations remain the same.

One convenient feature of the linearized evolution is that it does not allow for the planar reorganization of the ion crystal.    
This means the mode amplitudes can be calculated directly from the positions and velocities of the ions, which may not always be the case for the full Coulomb evolution where reorganization is possible, and especially common for lower rotating wall frequencies. 
At lower rotating wall frequencies, the $\exb$ modes have lower frequencies, and possess correspondingly larger excursions from equilibrium, making reorganization of the ion crystal more likely.

\section{Mode initialization and amplitude calculation}

In our simulations, we can individually initialize all $3N$ modes of the ion crystal. 
This consists of prescribing the initial mode amplitudes and phases. 
Typically, we assign all modes belonging to a particular branch of the mode spectrum the same initial amplitude, which corresponds to a preset mode branch temperature.
The phases of all modes are then chosen randomly from a uniform distribution. 
A detailed description of this process is given in the appendix of Ref.~\cite{Tang2021}.

It is important to note that our mode initialization procedure may realize ion crystals with slightly different initial energies $\mathrm{KE}_\perp, \mathrm{KE}_\parallel, \text{ and } \mathrm{PE}$ than anticipated, as it relies on the assumption that the mode amplitudes are small, and the linear approximation is valid. 
For all mode amplitudes initialized to correspond to a temperature of 10 mK, we find that the difference from the expected energy for $\mathrm{KE}_\perp, \mathrm{KE}_\parallel \text{ and } \mathrm{PE}$ and their values calculated after initialization, are typically no more than 1 mK.
The biggest discrepancies are seen for $\mathrm{PE}$, which arise from nonlinear corrections in the energy from the Coulomb interaction between the ions. We note that for a given temperature, the $\exb$ modes will exhibit the largest amplitude excursions because of their low frequencies.  Therefore the linear approximation will breakdown first for the $\exb$ modes.

After the simulation is complete, and assuming that reconfiguration of the ion crystal has not occurred, we can calculate the mode amplitudes directly from the positions and velocities of the ions.  
If a reconfiguration of the ion crystal has occurred, then the displacements from the equilibrium positions of the original configuration will lead to erroneously large mode amplitudes, and the analysis fails.

The mode amplitude calculation is used to directly calculate $\text{T}_{\text{cyc}}, \text{T}_{\text{axl}}, \text{ and } \text{T}_{\text{ExB}}$ in Fig.~3 of the main text. 
To calculate the mode amplitudes, the ion crystal's position and velocity coordinates in the lab frame are transformed to the rotating frame. 
The transformation of the planar position coordinates can be done by applying a rotation matrix to the $x \text{ and } y$ coordinates of the ions, i.e.,  
\begin{equation}
\label{eqn:rotatingmatrix}
\begin{aligned}
\begin{bmatrix}
x' \\
y' \\
\end{bmatrix}
=
\begin{bmatrix}
\cos(\omega_r t) & -\sin(\omega_r t) \\
\sin(\omega_r t) & \cos(\omega_r t) \\
\end{bmatrix}
\begin{bmatrix}
x \\
y \\
\end{bmatrix},
\end{aligned}
\end{equation}
where $x' \text{ and } y'$ are the rotating frame coordinates, and $x \text{ and } y$ are the lab frame coordinates. 
By making the coordinate transformation to the rotating frame, displacements from equilibrium can be calculated. 
The velocity coordinates are similarly transformed into the rotating frame. 
With the rotating frame displacements and velocities, the mode amplitudes can be calculated by projecting the displacements and velocities onto the mode eigenvectors.  
This yields a corresponding set of mode amplitudes for the set of mode eigenvectors. 
Finally, the energy of each mode can be calculated from the mode amplitudes as described in Eq.~(11) of Ref. \cite{Shankar2020}.

However, it is essential to note, that even in the absence of crystal reorganization, the mode branch temperatures calculated exhibit errors indicative of the underlying linear approximation employed. 
For example, we observe in Fig.~3 of the main text that the nonlinear simulations show variations and spikes in $\text{T}_\text{ExB}$ near t = 0, which are almost certainly due to a breakdown of the linear approximation at $\text{T} =  10 \text{ mK}$. 
In Fig.~3(d), these errors diminish as $\text{T}_\text{ExB}$ decreases, aligning with the expectation that the linear approximation gains increased accuracy at lower temperatures.

\section{Long Time Laser Cooling Simulations}

\begin{figure}[h!]
    \includegraphics[width=\linewidth]{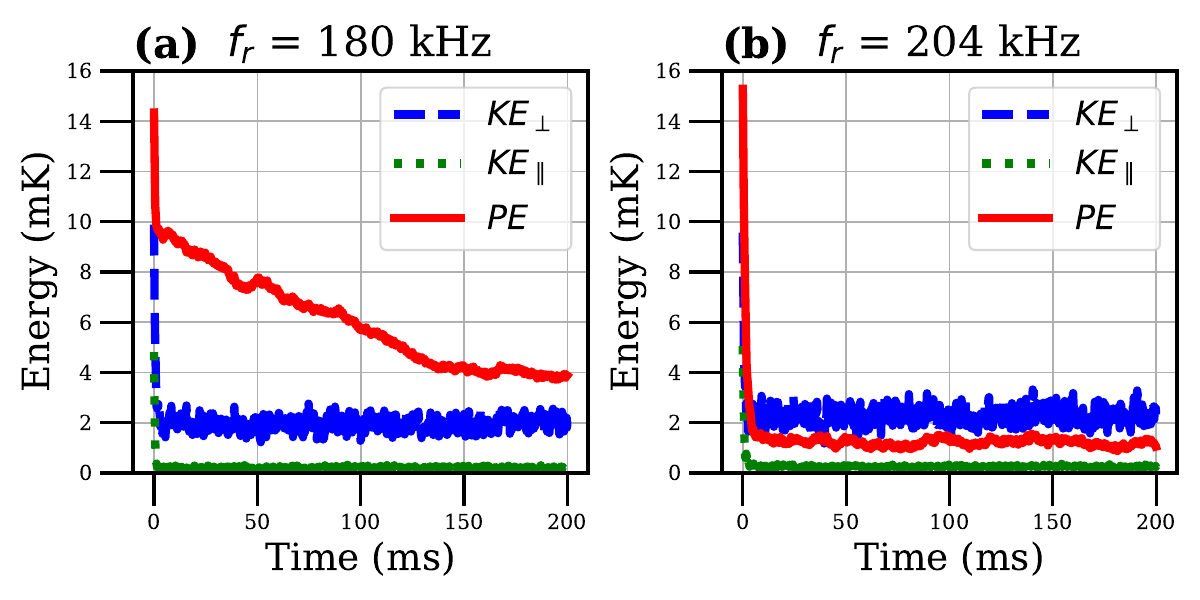} 
    \caption{\label{fig:longEvolution} 
    \textbf{Long Time Evolution of the Planar Ion Crystal.} 
    Two planar ion crystals with different rotating wall frequencies are shown.    
    Both are initialized with all mode amplitudes corresponding to $T = 10 \text{ mK}$, and laser cooled for 200 ms.    
    \textbf{(a)} Time evolution of energies for the ion crystal rotating at $\omega_r/(2\pi) 180 \text{ kHz}$. 
    The laser cooling simulation shows a slow decrease in $\text{PE}$ over roughly 150 ms, after which $\text{PE}$ achieves a minimum value around 4 mK.  
    The other energies, $\text{KE}_\perp \text{ and } \text{KE}_\parallel$, cool near instantaneously on this long timescale. 
    The initial rapid cooling of $\text{PE}$ is due to a decrease in drumhead axial mode energies. 
    This decrease in $\text{PE}$ corresponds to the decrease seen in $\text{KE}_\parallel$ during the start of the simulation. 
    \textbf{(b)} Time evolution of energies for the ion crystal with rotating wall frequency near $\omega_{\text{crit.}}$, $\omega_r = 2\pi \times 204 \text{ kHz}$. 
    The laser cooling simulation shows a rapid decrease in $\text{PE}$ which stabilizes to roughly $\text{PE} \sim 1 \text{ mK}$ for the remainder of the evolution. 
    The cooling of $\text{KE}_\perp \text{ and } \text{KE}_\parallel$ appears near-identical in both simulations, suggesting that changing $\omega_r$ strongly affects the cooling dynamics of $\text{PE}$ in particular. 
    }
\end{figure}    

In this section, we describe long time laser cooling simulations ($> 10 \text{ ms}$). 
The results are shown in Fig.~\ref{fig:longEvolution}.
We initialize the same ion crystals discussed in Fig.~2 of the main text.
The ion crystals' equilibrium configurations are shown in Fig.~\ref{fig:crystal}. 
These ion crystals are evolved for 200 ms with the same laser cooling parameters described in Fig.~1 of the main text.

First we consider the long time evolution of the crystal simulated in Fig.~2(c) of the main text, shown here in Fig.~\ref{fig:longEvolution}(a). 
The kinetic energies, $\text{KE}_\perp \text{ and } \text{KE}_\parallel$, are nearly instantaneously cooled on this long timescale, however, $\text{PE}$ cools on a timescale of hundreds of milliseconds. 
After a period of $150 \text{ ms}$, $\text{PE}$ achieves a value of roughly 4 mK. 
This is much higher than the mean values of roughly 2 mK for $\text{KE}_\perp$ and 0.5 mK for $\text{KE}_\parallel$. 
For the remainder of the laser cooling simulation ($> 150 \text{ ms}$) , $\text{PE}$ appears to fluctuate around a value of 4 mK.

Next we consider the long time evolution of the crystal described in Fig.~2(d) of the main text and shown here in Fig.~\ref{fig:longEvolution}(b). 
As described in the main text, for this value of $\omega_r$, many of the lower frequency drumhead modes are near resonant with the $\exb$ modes. 
Cooling of the kinetic energies, $\text{KE}_\perp \text{ and } \text{KE}_\parallel$, occurs rapidly.
However, $\text{PE}$ cooling also occurs on a very short timescale compared to the full integration time of $200 \text{ ms}$.  
After the initial several milliseconds of cooling, all energies fluctuate around a mean value for the remainder of the laser cooling process. 
This suggests that, with $\omega_r$ near $\omega_{\text{crit.}}$, all energies can be rapidly cooled. 
Furthermore, the cooling appears to achieve a lower steady-state energy for $\text{PE}$ of about 1 mK as compared to the 4 mK  observed for the crystal rotating at $\omega_r/(2\pi)=180\text{ kHz}$, shown in Fig.~2(a).

In real experiments, the slow cooling of $\text{PE}$ shown in Fig.~\ref{fig:longEvolution}(a) may not be sufficient to overcome external heating rates. 
Therefore, to guarantee sufficient cooling of in-plane motions, it may be vital to couple axial and planar motions by increasing $\omega_r$, as done in Fig.~2(b).
With this technique, the lowest attainable steady-state energy for $\text{PE}$ also seems to be lower.

\bibliography{bib}{}
\bibliographystyle{apsrev4-2}